\documentclass{article}
\usepackage{spconf,amsmath,graphicx}
\usepackage{amsfonts}
\usepackage{enumitem}
\setlist{nosep, leftmargin=14pt}
\usepackage{color}
\usepackage{mwe} % to get dummy images

\title{Reconstruction and segmentation of parallel MR data using image domain Deep-SLR}
\name{Aniket Pramanik, Mathews Jacob\thanks{This work is supported by NIH 1R01EB019961-01A1.}}
\address{The University of Iowa, Iowa City, USA}
\begin{document}
%\ninept
%
\maketitle
\begin{abstract}
	The main focus of this work is a novel framework for the joint reconstruction and segmentation of parallel MRI (PMRI) brain data. We introduce an image domain deep network for calibrationless recovery of undersampled PMRI data. The proposed approach is the deep-learning (DL) based generalization of local low-rank based approaches for uncalibrated PMRI recovery including CLEAR \cite{trzasko2012clear}. Since the image domain approach exploits additional annihilation relations compared to k-space based approaches, we expect it to offer improved performance. To minimize segmentation errors resulting from undersampling artifacts, we combined the proposed scheme with a segmentation network and trained it in an end-to-end fashion. In addition to reducing segmentation errors, this approach also offers improved reconstruction performance by reducing overfitting; the reconstructed images exhibit reduced blurring and sharper edges than independently trained reconstruction network.
\end{abstract}
\section{Introduction}
\label{sec:intro}

The segmentation of MR images is vital for the quantification of disease progression. For instance, the atrophy of important brain regions (e.g. hippocampal sub-regions, cortical thickness) are established biomarkers for progression in Alzhiemers disease; the volume estimates of these regions play important roles in the early diagnosis and prognosis of dementia subjects. Several segmentation methods, including classical  k-means clustering algorithms, deformable templates, and state of the art convolutional neural networks are available. These methods exploit the coherence of image intensities within similar tissues (e.g. gray matter, white matter) as well as edges between tissue boundaries to obtain good segmentation. Clearly, the ability to accurately resolve the small sub-regions depend on the spatial resolution of the images.

A challenge with the acquisition of high resolution MRI data is the long scan time, which is especially challenging for older adults. Long acquisition times are also associated with extensive motion artifacts. Modern MRI methods often rely on acceleration methods including PMRI and compressive sensing to reduce the scan time. Calibrated methods such as SENSE/GRAPPA \cite{pruessmann1999sense, griswold2002generalized} as well as recent calibrationless structured low-rank methods \cite{uecker2014espirit,jacob2020structured} have been introduced to recover the images from undersampled measurements. Our recent work has shown that  linear relations between their k-space measurements can be capitalized using k-space DL strategies \cite{pramanik2020deep},which are more computationally efficient than classical methods. Despite the great progress made in image reconstruction, undersampling artifacts and blurring of image edges are inevitable at high acceleration factors; these artifacts can deteriorate the performance of segmentation algorithms that exploit the edges and the coherence of image intensities within regions that would be impacted by undersampling. 

We introduce a novel framework for deep-learning based calibration-free MRI reconstruction and segmentation. The main contributions of this work are (1) the development of a novel image domain deep structured low-rank framework for calibration-free PMRI, motivated by locally low-rank methods used for PMRI \cite{trzasko2012clear} 
% http://archive.ismrm.org/2012/0517.html
and (2) the development of a joint segmentation-reconstruction framework to minimize segmentation errors introduced by undersampling artifacts and to improve reconstruction quality. The CLEAR formulation exploits the low-rank structure of image patches from sensitivity weighted images \cite{trzasko2012clear} . The low-rank structure results in inter and intra patch annihilation relations on the sensitivity weighted images. The annihilation relations vary spatially, depending on the coil sensitivities. An iterative reweighted formulation of the nuclear norm minimization algorithm in CLEAR as in \cite{pramanik2020deep} results in an alternating scheme; the algorithm alternates between data consistency steps and \emph{denoising/projection} using a spatially varying filterbank. Motivated by \cite{pramanik2020deep}, we propose to replace the filterbank with an  image domain CNN module \cite{pramanik2020deep}. In this work, we propose to use a UNET for the image domain CNN. We pre-learn the parameters of the unrolled algorithm, where the CNN parameters are shared across iterations, from exemplar data. The main distinction of this work with \cite{pramanik2020deep} is the formulation in the image domain; the increased number of annihilation relations between multi-channel images in the image domain compared to k-space that is exploited by \cite{griswold2002generalized,uecker2014espirit,jacob2020structured,pramanik2020deep} translate to improved performance.

To reduce the sensitivity of the segmentation algorithm to undersampling artifacts, we consider the end-to-end training of a cascade of the proposed reconstruction network with a segmentation network. We use a loss-metric, which is the sum of the mean square image reconstruction error and segmentation error to train the cascade network. The end-to-end training is expected to reduce segmentation errors compared to the straightforward cascade of individual algorithm. Because the loss metric is a combination of reconstruction and segmentation errors, one would expect the quality of the training images recovered by the reconstruction network to be inferior to the one trained with only the reconstruction loss. However, we conjecture that the segmentation loss term further reduces the generalization error, thus improving the mean square error on the test data. Specifically, the end-to-end optimization strategy ensures that the reconstructed images preserve edges and exhibit good coherence between regions with similar anatomical properties, and preserve edges, which will ensure good segmentation. In this work, we propose to use a UNET for segmentation; as in the case with the reconstruction, any segmentation network can be used within the proposed framework. In this work, we consider a simple segmentation setting, where we consider the segmentation of MR images to gray matter (GM), white matter (WM), and cerebrospinal fluid (CSF) regions.
\begin{figure*}
	\centering
	\includegraphics[scale=0.4,keepaspectratio=true,trim={3.2cm 5.0cm 3.0cm 5.2cm},clip]{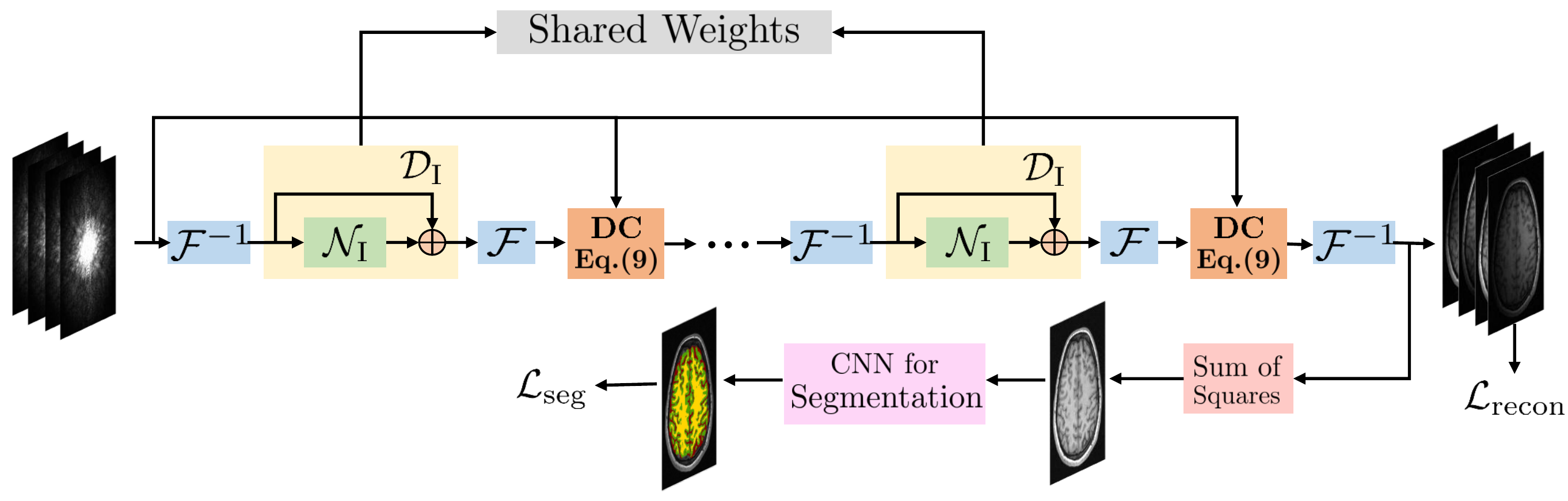}
	\caption{Proposed I-DSLR-SEG-E2E network architecture. A K-iteration I-DSLR network is cascaded with a CNN for segmentation. It is trained end-to-end.}
	\label{fig:casc_arch}
	%\vspace{-2em}
\end{figure*}
\section{Methods}
\label{bg}
\subsection{Image Domain Deep-SLR for PMRI}
The forward model for PMRI recovery can be defined as, 
\begin{equation}
\label{fmodel}
\mathbf b_i = \underbrace{\mathcal S \mathcal F}_{\mathcal A} ~\gamma_i + \mathbf n_i, \hspace{2pt} i = 1 \ldots N
\end{equation}
where $\mathbf b_i$ are noisy Fourier coefficients of $i^{th}$ coil image $\gamma_i$ corrupted by Gaussian noise $\mathbf n_i$, $N$ is the total number of coils, $\mathcal F$ is the Fourier operator and $\mathcal S$ denotes the sampling operator. 
The CLEAR formulation makes the locally-low rank assumption; it assumes that matrix composed of patches of $\gamma_i(\mathbf s)$ are low-rank. Specifically, if we assume $P_{\mathbf s_0}$ to be a patch extraction operator that extracts $M\times M$ patches centered at $\mathbf s_0$, the matrices
\begin{equation}\label{key}
\boldsymbol\Gamma_{\mathbf s} = \begin{bmatrix}
P_{\mathbf s}(\gamma_1)| & \ldots & |P_{\mathbf s}(\gamma_N)
\end{bmatrix}
\end{equation}  
are low-rank. The CLEAR approach solves for $\Gamma = \begin{bmatrix}
\gamma_1,.. ,\gamma_N
\end{bmatrix}$ as the nuclear norm minimization problem 
\begin{equation}\label{clear}
\boldsymbol \Gamma = \arg \min_{\boldsymbol \Gamma} \|\mathcal A(\boldsymbol\Gamma)-\mathbf B\|^2 + \lambda\sum_s \|\boldsymbol\Gamma_{\mathbf s}\|_*
\end{equation}

If we assume the rank of $\boldsymbol \Gamma_{\mathbf s}$ to be $r<N$, this implies that there exists $N-r$ null-space vectors $\mathbf u_{\mathbf s,j}$ and $\mathbf v_{\mathbf s,j}$ such that $\mathbf u_{\mathbf s,j}\boldsymbol \Gamma_{\mathbf s}=0$ and $\boldsymbol \Gamma_{\mathbf s}\mathbf v_{\mathbf s,j}=0$. %We thus have 
%\begin{equation}
%\label{left}
%\left\langle u_{\mathbf s,j}, \gamma_i\right \rangle = 0; ~~\forall i=1,..,M; j=1,..N-r
%\end{equation}
%Likewise, we also have 
%\begin{equation}
%\label{right}
%\sum_i v_{\mathbf s,j} \mathbf P_{\mathbf s}(\gamma_i) = 0; ~~ \forall j=1,..,N-r
%\end{equation}
If we consider the vertical concatenation of the patches denoted by the vector $\mathbf p_{\mathbf s}$, we can express the above relations compactly as 
\begin{equation}\label{mtxform}
\underbrace{\begin{bmatrix}
\mathbf u_{\mathbf s,j}&\ldots&0\\
\vdots&\vdots&\vdots\\
0&0&\mathbf u_{\mathbf s,j}\\
\hline\\
v_{\mathbf s,j}(1)\mathbf e_0&\ldots&v_{\mathbf s,j}(M^2)\mathbf e_0\\
\vdots&\vdots&\vdots\\
v_{\mathbf s,j}(1)\mathbf e_{M^2}&\ldots&v_{\mathbf s,j}(M^2)\mathbf e_{M^2}\\
\end{bmatrix}}_{\mathbf Q_{\mathbf s}}
\underbrace{\begin{bmatrix}P_{\mathbf s}(\gamma_1)\\
\vdots\\
P_{\mathbf s}(\gamma_N)\\
\end{bmatrix}}_{\mathbf p_{\mathbf s}} = \boldsymbol 0
\end{equation}
Here, $\mathbf e_i$ denotes the canonical basis vectors of the patches. We note that the inner-products of the patches with the $M^2+N$ row vectors in \eqref{mtxform} can be viewed as the convolution of the  multichannel volume $\boldsymbol\Gamma$ by flipped filters of size $M\times M\times N$, evaluated at $\mathbf s$. Hence, \eqref{mtxform} can be compactly expressed as 
\begin{equation}\label{convform}
\left(\mathbf q_s * \boldsymbol\Gamma\right)(\boldsymbol s) = \boldsymbol 0,
\end{equation}
where $\left(\mathbf q_s * \boldsymbol\Gamma\right)(\mathbf s)$denotes the output of a multichannel filter bank $\mathbf q_s$ evaluated at $\mathbf s$. We note that the image domain formulation has more annihilation relations than the corresponding k-space approaches \cite{griswold2002generalized,uecker2014espirit,jacob2020structured,pramanik2020deep}. Specifically, any local filter than annihilates $\rho$ (e.g wavelet filters that vanish in smooth regions) will annihilate all of the multichannel patches. We expect the ability of the formulation to exploit intra and inter channel annihilation relations to translate to improved performance.

Using the iterative reweighted formulation in \cite{pramanik2020deep} to minimize the nuclear norm minimization problem \eqref{clear}, we alternate between 
\begin{equation}\label{matrixform}
\boldsymbol \Gamma = \arg \min_{\boldsymbol \Gamma} \|\mathcal A(\boldsymbol\Gamma)-\mathbf B\|^2 + \lambda\sum_s \left\|(\mathbf q_s * \boldsymbol\Gamma)(\boldsymbol s)\right\|^2,
\end{equation}
and the derivation of the $\mathbf q_s$ matrices as in \cite{pramanik2020deep,jacob2020structured}. We note that $\mathbf q_s$ is a spatially varying filterbank that is derived from the signal patches $\boldsymbol \Gamma_{\mathbf s}$ itself. Motivated by \cite{pramanik2020deep}, we replace the spatially varying and signal-dependent filterbank by a deep CNN; we note that CNNs can closely approximate spatial variations in the linear filterbank structure. The proposed algorithm is formulated as
\begin{equation}\label{dlform}
\boldsymbol \Gamma = \arg \min_{\boldsymbol \Gamma} \|\mathcal A(\boldsymbol\Gamma)-\mathbf B\|^2 + \lambda \|\mathcal N_{\rm I}(\boldsymbol \Gamma)\|^2,
\end{equation}
where $\mathcal N_{\rm I} = \mathcal I - \mathcal D_{\rm I}$ denotes a residual multichannel CNN; the input to the filterbank has $N$ channels corresponding to the coil sensitivity weighted images. Here, $\mathcal D_{\rm I}$ is a spatial domain CNN. We choose $\mathcal D_{\rm I}$ as a 2D spatial domain 
 \begin{eqnarray}
\widehat{\mathbf X}_{n} &=& \mathcal D_{\rm I}(\widehat{\boldsymbol \Gamma}_n)\\
\widehat{\boldsymbol \Gamma}_{n+1} &=& (\mathcal A^H  \mathcal A + \lambda \mathbf I)^{-1}(\mathcal A^H \mathbf B + \lambda \widehat{\mathbf X}_{n})
\end{eqnarray}   
We propose to learn the parameters of the CNN in the unrolled algorithm from exemplar data using an end-to-end optimization strategy. The loss of the supervised training is chosen as $\mathcal L_{\rm recon} = \|\boldsymbol \Gamma - \boldsymbol \Gamma_{gs} \|^2$, where $\boldsymbol\Gamma$ is the output of the unrolled CNN and $\boldsymbol \Gamma_{gs}$ is the gold standard multicoil data obtained from fully sampled measurements.  The Adam optimizer is used to minimize training loss at a learning rate of $10^{-4}$ for all the experiments. It is named as Image Domain Deep-SLR (I-DSLR). 

I-DSLR eliminates the need for calibration data for estimating the coil sensitivities or linear filters $\mathbf Q_s$. Moreover, this approach eliminates the need for singular value decompositions at each iterations that contributes heavily to the computational complexity of CLEAR \cite{trzasko2012clear}. Similar to MoDL \cite{aggarwal2018modl} and \cite{pramanik2020deep}, we share the parameters of the CNN block across iterations.
\section{Joint Reconstruction \& Segmentation}
\label{proposed}
As discussed previously, the straightforward cascade of reconstruction and segmentation algorithms can result in the propagation of error, thus downgrading segmentation quality. Specifically, the residual alias artifacts as well as blurring caused by undersampling can result in segmentation errors. To minimize this issue, we propose a multi-task deep network as shown in Fig. \ref{fig:casc_arch}. The reconstruction network is I-DSLR (described in Section \ref{bg}) with shared weights across iterations. A segmentation network is attached to the final iteration of the I-DSLR. The combined network is trained end-to-end. 

We use a weighted linear combination of normalized mean squared error $\mathcal L_{\rm recon}$ and pixel-wise multi-label cross entropy $\mathcal L_{\rm seg} = -\sum_p \boldsymbol{\phi}_{p}^{gs} \ln \boldsymbol{\phi}_{p}$ error for training. For each pixel $p$, $\boldsymbol{\phi}_{gs}$ is the gold standard segmentation on the sum-of-squares image obtained from $\boldsymbol{\Gamma}_{gs}$ and $\boldsymbol{\phi}$ is the segmentation CNN output. 
\begin{equation}\label{key}
\mathcal L_{\rm total} = \mathcal L_{\rm recon} + \beta \mathcal L_{\rm seg}
\end{equation}
The proposed multi-task network is initialized with weights obtained from pre-trained reconstruction and segmentation networks for training. Specifically, the segmentation UNET is pre-trained with fully sampled images. The reconstruction network is trained with undersampled k-space measurements. The pre-training for both the tasks were done with the brain images described in section \ref{exp_res}. $\beta  = 1$ is chosen to equally weigh both the losses.
\begin{figure*}
	\centering
	\includegraphics[scale=0.55,keepaspectratio=true,trim={1.8cm 6.5cm 1.4cm 6.8cm},clip]{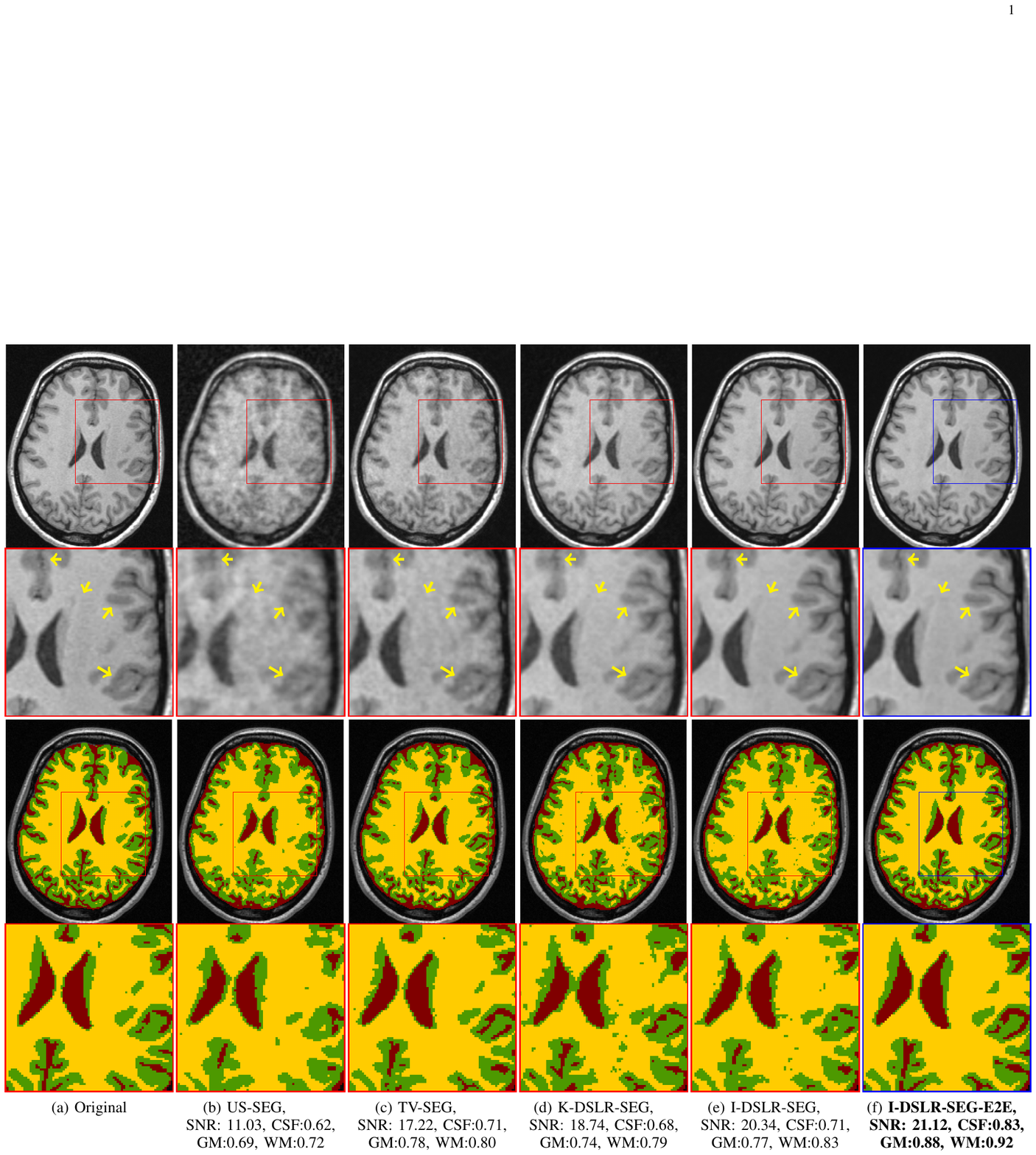}
	\caption{Comparison of reconstruction and segmentation quality of various methods on 6-fold undersampled k-space measurements. Reconstruction SNR in dB along with dice coefficients for CSF, GM and WM are reported for the particular slice. The methods in red box typically cascade separately trained tasks and the blue one is the proposed end-to-end training approach.}
	\label{fig:recon_seg}
\end{figure*}        
\section{Experiments and Results}
\label{exp_res}
We perform experiments on the publicly available Calgary Campinas Dataset (CCP) \cite{souza2018open}. It consists of 12-channel raw k-space data of T1-weighted brain MRI scans from a Discovery MR750 3T scanner for 67 subjects. The slice dimensions are 208 x 170 for axial view of the brain. Ground truth segmentations for the slices were generated using the FAST software which uses the standard k-means clustering technique for segmentation. Forty subjects (40 x 256 = 10240 slices) were used for training, 7 for validation and the remaining 20 for testing purposes. 2D non-uniform cartesian variable density undersampling masks with different acceleration factors were used for experiments; readout direction is orthogonal to axial slices.  

The proposed image domain Deep-SLR approach with segmentation algorithms pre-trained using fully sampled data (I-DSLR-SEG) is compared against k-space Deep-SLR method(K-DSLR-SEG), total variation (TV-SEG) and undersampled (US-SEG) in the same setting. We also show the benefit of end-to-end training by comparing the above methods against  I-DSLR-SEG-E2E, which is a cascade of I-DSLR and UNET Segmentation networks trained end-to-end (E2E). I-DSLR-SEG is the direct cascade of pre-trained I-DSLR and UNET-based segmentation networks. In TV-SEG, the segmentation network is trained and tested on images reconstructed using TV. Similarly, for US, the segmentation network is trained and tested on undersampled datasets.
 
A comparison of the methods is recorded and shown in Table \ref{tab:comp_arc} and Fig. \ref{fig:recon_seg} respectively. The I-DSLR reconstructions have sharper edges with more information preserved compared to US, TV and K-DSLR. The corresponding segmentation performance improves with increase in reconstruction quality. The end-to-end training strategy in I-DSLR-SEG-E2E further improves quality over I-DSLR-SEG. Its improved reconstruction can be attributed to the regularization by the segmentation network and vice-versa. I-DSLR-SEG-E2E alleviates errors propagated from reconstruction to segmentation unlike K-DSLR-SEG and I-DSLR-SEG settings.        
\begin{table}[h!]
	\fontsize{6}{10}
	\selectfont
	\centering
	\renewcommand{\arraystretch}{0.8}
	\begin{tabular}{|c|c|ccc|}
		\hline
		\multicolumn{5}{|c|}{8-Fold Accelerated Reconstruction and Segmentation} \\ \hline
		Methods& SNR & Dice CSF & Dice GM & Dice WM \\ \hline
		US-SEG & 10.21 & 0.658 & 0.745 & 0.763 \\
		TV-SEG & 16.13 & 0.701 & 0.772 & 0.803 \\
		K-DSLR-SEG & 17.82 & 0.749 & 0.781 & 0.835 \\
		I-DSLR-SEG & 19.28 & 0.763 & 0.799 & 0.856 \\
		I-DSLR-SEG-E2E & \textbf{19.85} & \textbf{0.802} & \textbf{0.862} & \textbf{0.907} \\ 
		\hline
		\end{tabular}
	\caption{Quantitative comparisons of reconstruction (SNR in dB) and segmentation (dice coefficients) quality for different methods. The metrics are averaged over 20 subjects.}
	\label{tab:comp_arc} 
\end{table} 
\section{Conclusion}
We introduced a novel image domain model-based DL approach for calibrationless PMRI recovery. It is a non-linear extension of locally low rank methods for calibrationless parallel MRI. The experiments show that additional annihilation relations exploited by I-DSLR approach offers better performance over k-space approach K-DSLR \cite{pramanik2020deep}. We introduce a 
multi-task framework where I-DSLR is cascaded with a segmentation dedicated DL network which is trained end-to-end. The networks regularize each other, thereby reducing the errors caused from undersampling artifacts. Experiments show that the segmentation accuracy depends on the reconstruction quality. I-DSLR-SEG-E2E reduces the propagation of errors from reconstruction to segmentation, this outperforming methods that cascade independently trained networks.         

\vspace{-1em}
\section{Compliance with Ethical Standards}
\vspace{-1em}
This research study was conducted using a public domain  Calgary Campinas Dataset (CCP) \cite{souza2018open}.

\bibliographystyle{IEEEbib}
\bibliography{strings,refs}

\begin{thebibliography}{1}

\bibitem{pruessmann1999sense}
Pruessmann et~al.,
\newblock ``{SENSE: sensitivity encoding for fast MRI},''
\newblock {\em MRM}, vol. 42, no. 5, pp. 952--962, 1999.

\bibitem{griswold2002generalized}
Griswold et~al.,
\newblock ``{Generalized autocalibrating partially parallel acquisitions
  (GRAPPA)},''
\newblock {\em MRM}, vol. 47, no. 6, pp. 1202--1210, 2002.

\bibitem{uecker2014espirit}
Uecker et~al.,
\newblock ``{ESPIRiT—an eigenvalue approach to autocalibrating parallel MRI:
  where SENSE meets GRAPPA},''
\newblock {\em MRM}, vol. 71, no. 3, pp. 990--1001, 2014.

\bibitem{jacob2020structured}
Mathews et~al.,
\newblock ``{Structured Low-Rank Algorithms: Theory, Magnetic Resonance
  Applications, and Links to Machine Learning},''
\newblock {\em IEEE SPM}, vol. 37, no. 1, pp. 54--68, 2020.

\bibitem{pramanik2020deep}
Pramanik et~al.,
\newblock ``{Deep Generalization of Structured Low-Rank Algorithms
  (Deep-SLR)},''
\newblock {\em IEEE TMI}, 2020.

\bibitem{trzasko2012clear}
Joshua~D Trzasko and Armando Manduca,
\newblock ``{CLEAR: Calibration-free parallel imaging using locally low-rank
  encouraging reconstruction},''
\newblock in {\em ISMRM}, 2012, vol. 517.

\bibitem{aggarwal2018modl}
Aggarwal et~al.,
\newblock ``{MoDL: Model-based deep learning architecture for inverse
  problems},''
\newblock {\em IEEE TMI}, vol. 38, no. 2, pp. 394--405, 2018.

\bibitem{souza2018open}
Souza et~al.,
\newblock ``An open, multi-vendor, multi-field-strength brain mr dataset and
  analysis of publicly available skull stripping methods agreement,''
\newblock {\em NeuroImage}, vol. 170, pp. 482--494, 2018.

\end{thebibliography}

\end{document}